# PULSED NEUTRON SOURCE USING 100-MEV ELECTRON LINAC AT POHANG ACCELERATOR LABORATORY[*]

G. N. Kim[#], H. S. Kang, Y. S. Lee, M. H. Cho, I. S. Ko, and W. Namkung
Pohang Accelerator Laboratory, POSTECH, Pohang 790-784, Korea

*Abstract*

The Pohang Accelerator Laboratory operates an electron linac for a pulsed neutron source as one of the long-term nuclear R&D programs at the Korea Atomic Energy Research Institute. The designed beam parameters are as follows; The nominal beam energy is 100 MeV, the maximum beam power is 10 kW, and the beam current is varied from 300 mA to 5A depends on the pulse repetition rate. The linac has two operating modes: one for the short pulse mode with the repetition rates of 2 - 100 ns and the other for the long pulse mode with the 1 µs repetition rate. We tested an electron linac based on the existing equipment such as a SLAC-5045 klystron, two constant gradient accelerating sections, and a thermionic RF-gun. We investige the characteristics of the linac, and we report the status of the pulsed neutron source facility including a target system and the time-of-flight paths.

## 1 INTRODUCTION

The nuclear data project as one of the nation-wide nuclear R&D programs was launched by the KAERI in 1996 [1]. Its main goals are to establish a nuclear data system, to construct the infrastructure for the nuclear data productions and evaluations, and to develop a highly reliable nuclear data system. In order to build the infrastructure for the nuclear data production, KAERI is to build an intense pulsed neutron source by utilizing accelerator facilities, technologies, and manpower at the Pohang Accelerator Laboratory (PAL). The PAL proposed the Pohang Neutron Facility (PNF), which consists of a 100-MeV electron linac, a water-cooled Ta target, and at least three different time-of-flight (TOF) paths [2]. We designed a 100-MeV electron linac [3] and constructed an electron linac based on experiences obtained from construction and operation of the 2-GeV linac at PAL.

In this paper, we describe the characteristics of the electron linac, and then we present the status of the Pohang Neutron Facility (PNF).

___________________
[*]Work supported in part by POSCO, MOST, and KAERI
[#] Email: gnkim@postech.ac.kr ; Joint appointment at the Institute of High Energy Science, Kyungpook National Univ., Taegu , Korea.

## 2 CHARACTERISTICS OF E-LINAC

### 2.1 Construction of electron linac

We constructed an electron linac for the various R&D activities of the neutron facility by utilizing the existing components and infrastructures at PAL on December 1997 [4]. The design beam parameters of the electron linac are as follows; The beam energy is 100 MeV with the energy spread less than 1%, the peak current is 20A, the beam pulse width is 6-µs, and the rms normalized emittance is less than 30 $\pi$mm-mrad. The electron linac consists of a thermionic RF-gun, an alpha magnet, four quadrupole magnets, two SLAC-type accelerating structure, a quadrupole triplet, and a beam-analyzing magnet. A 2-m long drift space is added between the first and second accelerating structures to insert an energy compensation magnet or a beam transport magnet for the future FEL research. The RF-gun is a one-cell cavity with a tungsten dispenser cathode of 6 mm diameter. The RF-gun produced an electron beam with an average current of 300 mA, a pulse length of 6-µs, and aproximate energy of 1 MeV [5]. The measured rms emittance for the beam energy of 1 MeV was 2.1 $\pi$mm-mrad. The alpha magnet is used to match the longitudinal acceptance from the RF- gun to the first acceleration structure. Electrons move along a $\alpha$-shaped trajectory in the alpha magnet, and the bending angle is 278.6$^\circ$. The higher energy electron has a longer path length than the lower energy electron, thus the length of electron beams is not lengthened or is shortened in the beam transport line from the RF-gun to the first accelerating structure.

Four quadrupole magnets are used to focus the electron beams in the beam transport line from the thermionic RF-gun to the first accelerating structure. The quardupole triplet installed between the first and the second accelerating structure is used to focus the electron beam during the transport to the experimental beam line at the end of linac. There are three beam current transformers (BCT) and three beam profile monitors for beam instrumentation. The BCT is the toroidal shape of ferrite core 25 turns-wound by 0.3 mm diameter enameled wire. The beam-analyzing magnet has a bending angle of 30 degrees and zero pole-face rotation.

The main components of RF system consist of a SLAC 5045 klystron and an 80-MW modulator, and RF waveguide components, etc. Two branched waveguides through a 3 dB power divider from the main klystron waveguide are connected accelerating structures. A high power phase-shifter is inserted in the waveguide line of the second accelerating structure. The branched waveguide with a 10 dB power divider from the main waveguide is connected to the RF-gun cavity through a high-power phase-shifter/attenuator and a high-power circulator. The circulator is pressurized with dry nitrogen at 20 psig. Two waveguide windows isolate the circulator from the evacuated waveguide line.

*2.2 Beam Acceleration*

After the RF-conditioning of the accelerating structures and the wave-guide network, we tested the beam acceleration. The maximum RF power from a SLAC 5045 klystron was up to 45 MW. The RF power fed to the RF-gun was 3 MW. The maximum energy is 75 MeV, and the measured beam currents at the entrance of the first accelerating structure and at the end of linac are 100 mA and 40 mA, respectively. The length of electron beam pulses is 1.8 μs, and the pulse repetition rate is 12 Hz. The measured energy spread is ± 1% at its minimum. The energy spread was reduced by adjustment of the RF phase for the RF-gun and by optimization of the magnetic field for the alpha magnet.

## 3 STATUS OF NEUTRON FACILITY

The Pohang Neutron Facility consists of a 100-MeV electron linac, a photo-neutron target, and at least three different time-of-flight (TOF) paths.

*3.1 Photo-neutron Target*

High-energy electrons injected in the target produce gamma rays via bremsstrahlung, these gamma rays then generate neutrons via photonuclear reactions. We are considering tantalum rather than fissile material because the technology for handling and the target characteristics are well known [6]. The neutron yield depends sensitively on the materials and the target geometry. The target system is desgined using the MC simulation codes, EGS4 and MCNP4. The target system, 4.9-cm in diameter and 7.4-cm in length, is composed of ten sheets of Ta plates, and there is 0.15-cm water gap between them, in order to cool the target effectively [7]. The estimated flow rate of the cooling water is about 5 liters per minute in order to maintain below 45 °C. The target housing is made of Titanium. The conversion ratio obtained from the MCNP4 code from a 100-MeV electron to neutrons is 0.032. The neutron yield per kW beam power at the target is $2.0 \times 10^{12}$ n/sec, which is about 2.5% lower than the calculated value based on the Swanson's formula [8]. Based on this study, we constructed a water-cooled Ta-target system.

*3.2 Time-of-Flight Path*

The pulsed neutron facility based on the electron linac is a useful tool for high-resolution measurements of micro-scopic neutron cross-sections with the TOF method. In the TOF method, the energy resolution of neutrons depends on the TOF path length. Since we have to utilize the space and the infrastructures in the laboratory, the TOF paths and experimental halls are placed perpendicular to the electron linac. We constructed a 15 m long TOF path perpendicular to the electron linac. The TOF tubes were made by 15 and 20 cm diameter of the stainless pipes.

*3.3 Neutron Production Experiment*

The neutron production facility PNF was operated from January 2000. We measured the neutron TOF spectra for several samples with the pulsed neutron beam. The experimental arrangement for the neutron TOF spectrum measurement is shown in Fig. 1. The target is located in the position where the electron beam hits the center of the photoneutron target, which is also aligned with the center- line of the TOF tube. The target was installed at the center of a water modulator, 50 cm in diameter and 30 cm long aluminum cylindrical shape, to moderate the fast neutrons. A lead block, 20cm x 20cm x 10 cm, was inserted in front of the TOF tube to reduce the gamma-flash generated by the electron burst from the target. The sample was placed at the midpoint of the flight path. There is 1.8 m thick concrete between the target room and the detector room.

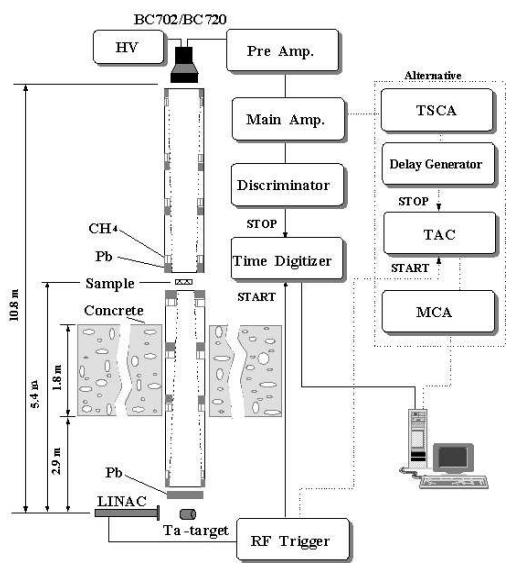

Fig. 1. Experimental arrangement for TOF spectra measurement

For the neutron TOF spectrum measurement, a $^6$Li glass scinitillator BC702, 12.5 cm in diameter and 1.5 cm in thickness was mounted on an EMI-93090 photomultiplier, and it is used as a neutron detector. In order to monitor the neutron intensity during the measurement, a $BF_3$ proportional counter, 1.6 cm in diameter and 5.8 cm long was placed inside the target room at a distance of about 6 m from the target. The BF$_3$ counter was inserted in a 30.5 cm diameter polyethylene sphere and surrounded by 5 cm thick borated polyethylene plates and Pb bricks to shield thermal neutrons and gamma flash.

During the experiment, the electron linac was operated with a repetition rate of 12 Hz, a pulse width of 1.5 μs, a peak current of 30 mA, and electron energy of 60 MeV. The total operated period is about two weeks from March 2000. During this period, we spend most of time to check the radiation level around the facility and to reduce the electronic noises originated from the RF source.

The block diagram of the data acquisition system is also shown in Fig. 1. The TOF signal from the $^6$Li-ZnS(Ag) scintillator was connected through an ORTEC-113 pre-amplifier to an ORTEC-571 amplifier, the amplifier output was fed into a discriminator input and used a stop signal of a 150 MHz time digitizer (Turbo MCS). The lower threshold level of the discriminator was set to 30 mV. The Turbo MCS was operated as a 16,384-channel time analyzer. The channel width of the time analyzer was set to 0.5 μs. The 12 Hz trigger signal for a modulator of an electron linac was connected to an ORTEC-550 single channel analyzer (SCA), the output signal was used as a start signal of a Turbo MCS. The Turbo MCS is connected to a personal computer. The data were collected, stored and analyzed on this computer. In order to determine the flight path distance for our facility, we used neutron TOF spectra for Sm, Ta, W, and Ag sample runs. A Cd filter of 0.5 mm in thickness was used to suppress thermal neutrons. The samples were placed at the midpoint of the flight path.

The resonance energy $E$ and the channel number $I$ in Table 2 are used to find the flight path length $L$ in the following equation by the method of least squares fitting.

$$I = \frac{72.3 \times L}{\Delta t \times \sqrt{E}} + \frac{\tau}{\Delta t}$$

In the above equation, $\Delta t$ is the channel width of the time digitizer and set to 0.5 μs. The delay time $\tau$ is the time difference between the start signal from the RF trigger and the real zero time. The flight path length L is determined from the fitting. As shown in Fig. 2, the results of the fit are: L=10.81±0.02 m and τ=0.87 μs.

## 4. SUMMARY

The nuclear data project was launched by KAERI from 1996 in order to support nuclear R&D programs, medical and industrial applications. We have constructed and tested an electron linac for the pulsed neutron facility by utilizing the existing components and infrastructures at PAL. The characteristics of accelerated electron beams are about 75 MeV of energy, 12 Hz of repetition rate, 1.8μs of pulse width and about 40 mA of peak current. We made a 15-m TOF path perpendicular to the linac in order to test a Ta-target system and a data acquisition system. We tested the neutron production and measured the neutron TOF spectra. Using the resonance energy of TOF spectra for various samples, we measured the TOF path length of the neutron facility.

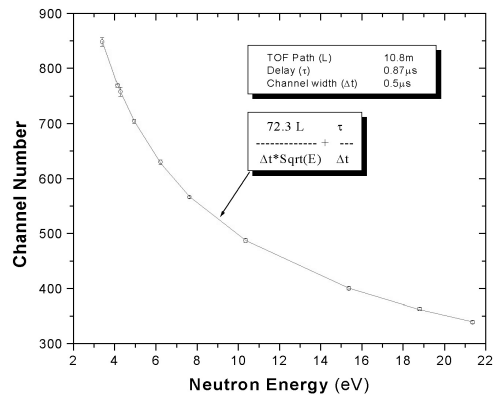

Fig. 2 A fit of the flight path length to resonance energies